%%% THIS IS A Plain TeX file.
%%% This is a Plain TeX file.
\magnification=1200

%%\baselineskip=24pt
%%\hfill {IUT-PHYS }
%%%% \hfill {September 2002} \vskip 0.1in

\font\bigbf=cmbx10  scaled\magstep2 \vskip 0.2in
\centerline{\bigbf The Klein-Gordon and the Dirac oscillators }
\vskip 0.1in \centerline{\bigbf in a noncommutative space }

\vskip 0.4in \font\bigtenrm=cmr10 scaled\magstep1
\centerline{\bigtenrm  B. Mirza$^{\ast}$ and M. Mohadesi } \vskip
0.2in

\centerline{\sl Department of Physics, Isfahan University of
Technology, Isfahan 84154, Iran } \vskip 0.1in
%% \centerline{\sl $
%%%^{\ddag}$Institute for Studies in Theoretical Physics and
%%%Mathematics, } \centerline{\sl P.O.Box 5746, Tehran, 19395, Iran}
\vskip 0.1in

%%\centerline{\sl Isfahan University of Technology,}
%%\centerline{\sl Isfahan 80971, Iran,}
\centerline{\sl $^{\ast}$E-mail: b.mirza@cc.iut.ac.ir}

\vskip 0.2in \centerline{\bf ABSTRACT} \vskip 0.1in

We study the Dirac and the klein-Gordon oscillators in a
noncommutative space. It is shown that the Klein-Gordon oscillator
in a noncommutative space has a similar behaviour to the dynamics
of a particle in a commutative space and in a constant magnetic
field. The Dirac oscillator in a noncommutative space has a
similar equation to the equation of motion for a relativistic
fermion in a commutative space and in a magnetic field, however a
new exotic term appears, which implies that a charged fermion in a
noncommutative space has an electric dipole moment.



\vfill\eject

\vskip 1in

\centerline{I. \bf \ Introduction } \vskip 0.1in

In the last few years theories in noncommutative space have been
studied extensively (for a review see Ref. [1]). Noncommutative
field theories are related to M-theory compactification [2],
string theory in nontrivial backgrounds [3] and quantum Hall
effect [4]. Inclusion of noncommutativity in quantum field theory
can be achieved in two different ways: via Moyal $\star$-product
on the space of ordinary functions, or defining the field theory
on a coordinate operator space which is intrinsically
noncommutative[1,5]. The equivalence between the two approaches
has been nicely described in Ref. [6]. A simple insight on the
role of noncommutativity in field theory can be obtained by
studying the one particle sector, which prompted an interest in
the study of noncommutative quantum mechanics (NCQM)
[7,8,9,10,11,12,13,14]. In these studies some attention was paid
to two-dimensional NCQM and its relation to the Landau problem.
It has been shown that the equation of motion of a harmonic
oscillator in a noncommutative space is similar to the equation
of motion of a particle in a constant magnetic field and in the
lowest  Landau level [13]. We generalize these relations to the
relativistic quantum mechanics. In particular, it is shown that
the Dirac and Klein-Gordon oscillators in a noncommutative space
have similar behaviour to the Landau problem in a commutative
space although an exact map does not exist. However, for the
Dirac oscillator there is a
 new term which is spin dependent. The noncommutative spaces can
 be realized as spaces where coordinate operator $ \hat x^\mu $
 satisfies the commutation relations

$$ [ \hat x^\mu , \hat x^\nu ] = i \theta^{\mu \nu } \eqno(1) $$

\noindent where $ \theta^{\mu \nu} $ is an antisymmetric tensor
and is of space dimension (length$)^2 $. We note that space-time
noncommutativity, $\theta^{0 i}\neq 0 $, may lead to some
problems with unitarity and causality. Such problems do not occur
 for the quantum mechanics on a noncommutative space with a usual
 commutative time coordinate. The noncommutative models specified
 by Eq. (1) can be realized in terms of a $\star$-product: the
 commutative algebra of functions with the usual product
 f(x)g(x) is replaced by the $\star$-product Moyal algebra:

 $$( f\star g)(x)= exp \ [{i\over 2} \theta_{\mu \nu
 }\partial_{x_\mu}\partial_{y_{\nu}}]\ f(x)g(y)|_{x=y} \eqno(2)$$
\noindent  In the case when $ [\hat p_i,\hat p_j]=0 $, the
noncommutative quantum mechanics

$$ H(\hat p,\hat x) \star \psi(\hat x)=E \psi(\hat x)\eqno(3) $$

\noindent reduces to the usual one described by [7-14],

$$ H(\hat p , \hat x) \psi ( x) = E \psi (x)\eqno(4)$$

\noindent where

$$ \hat x_i =x_i - {1\over 2 \hbar } \theta_{ij}p_j \ \ \ \ , \  \ \ \  \hat
p_i= p_i   \eqno(5) $$

\noindent The new variables satisfy the usual canonical
commutation relations:

$$ [x_i,x_j]=0 \ , \ [p_i, p_j]=0 \ , \ [x_i,p_j]=i \hbar \delta_{ij} \eqno(6)$$

\noindent This paper is organized as follows: in section 2, the
Klein-Gordon oscillator in a noncommutative space is investigated
and its map to the Landau problem in a commutative space is
given. In section 3, the Dirac oscillator in a noncommutative
space is defined and its relation to the Landau problem is
clarified.

\vskip 0.2in \centerline{II. \bf \ The klein-Gordon oscillator in
a noncommutative space } \vskip 0.1in

It is known that the nonrelativistic harmonic oscillator in a
noncommutative space has a similar behaviour to the Landau
problem in a commuatative space [10,11,13,14,20]. In this section
we investigate this relation in the relativistic case. The
klein-Gordon oscillator in a two dimensional (2+1 dimensional
space-time) commutative space can be defined by the following
equation

$$ c^2 (\vec p +imw\vec r)\cdot (\vec p- i m w \vec r)\psi =(E^2 -
m^2 c^4) \psi \eqno(7)$$

\noindent and the energy eigenvalues are given by

$$ E_{n_x n_y}^2=  2 m c^2 \hbar w (n_x + n_y + 1)+ m^2 c^4 -2mc^2 \hbar w  \eqno(8) $$

\noindent The nonrelativistic limit of the Eq. (7) is given by

$$ [{p^2\over {2m}} + {1\over 2} m w^2 r^2
]\psi =(\varepsilon + \hbar w) \psi \ , \  \ \ \ \ \ \
\varepsilon=E-mc^2 \eqno(9)$$

\noindent which justifies the name given to it. In a
noncommutative space one may describe the Klein-Gordon oscillator
by the following equation

$$ c^2  [(\vec p +imw\vec r)\cdot (\vec p - i m w \vec r)] \star \psi =(E^2 -
m^2 c^4) \psi \eqno(10)$$

\noindent which is equivalent to the equation below in a
commutative space ($\theta_{ij}=\epsilon_{ijk}\theta_k$)

$$ c^2 [\ \vec p + imw(\vec r + {{\vec \theta \times \vec p}\over {2 \hbar}})] \cdot
[ \ \vec p - i m w (\vec r + {{\vec \theta \times \vec p}\over {2
\hbar}})]\  \psi =(E^2 - m^2 c^4) \psi \eqno(11)$$

\noindent If we put $\theta_3 =\theta $ and choose the rest of the
$\theta$-components equal to zero (which can be done by a rotation
or a redefinition of coordinates), after straightforward
calculation, we arrive at the  following equation which can be
solved exactly

$$ c^2 [ (1+ {{m^2 w^2 \theta^2 }\over 4 \hbar^2})(p_x^2+p_y^2) + m^2 w^2 ( x^2 + y^2 )
 - {{m^2 w^2 \theta}\over \hbar}
L_z  ] \psi =(E^2 - m^2 c^4 + 2 m c^2\hbar w )\psi \eqno(12) $$

\noindent and the energy eigenvalues are given by

$$ E_{n_x n_y m_\ell}^2= 2 m c^2 \hbar w_1 (n_x+n_y +1 )- ({m^2 w^2 c^2 \theta \over
\hbar}) m_\ell \hbar +  m^2 c^4 - 2 m c^2 \hbar w   \eqno(13) $$

\noindent where

$$ w_1 = w {\sqrt{1+ {{m^2 w^2 \theta^2} \over {4 \hbar^2}}}}  \eqno(14) $$

 \noindent The energy eigenvalues indicate a similarity to the {\it normal Zeeman effect}. The
 nonrelativistic limit
of Eq. (12) is given by

$$  [ ({1\over 2 m} + {{m w^2 \theta^2 }\over 8 \hbar^2})(p_x^2+p_y^2) +
 {m w^2 (x^2+y^2) \over 2} - {{m w^2 \theta}\over  2 \hbar}
L_z  ] \psi =(\varepsilon+\hbar w) \psi \ , \ \ \
\varepsilon=E-mc^2 \eqno(15) $$

 \noindent The Klein-Gordon equation for a particle in a commutative
space and in a constant magnetic field can be written as

$$ c^2 [(\vec p- {e\over c}\vec A)] \cdot [(\vec p- {e\over c}\vec A)]
\psi= (W^2 - m^2 c^4) \psi  \eqno(16) $$

\noindent where

$$ A={{\vec B \times \vec r}\over 2} \eqno(17) $$

\noindent A straightforward calculation in the Coulomb gauge
yields

$$ c^2 [( p^2_x+p^2_y) + ({{{e^2 B^2}\over 4 c^2 }})(x^2+y^2) -{{e B}\over c} L_z]
\psi = (W^2 - m^2 c^4 )\psi \eqno(18) $$

\noindent  which is quite similar to Eq. (12), although an
invertible map between the coefficients does not exist. These
arguments can be extended to the three dimensional space and
similar results can be obtained. The Klein-Gordon oscillator in a
three dimensional (3+1) commutative space can be defined by the
following equation

$$ c^2 (\vec p +imw\vec r)\cdot (\vec p- i m w \vec r)\psi =(E^2 -
m^2 c^4) \psi  \eqno(19)$$

\noindent and the energy eigenvalues are given by

$$ E_{n_x n_y n_z}^2=  2 m c^2 \hbar w (n_x + n_y + n_z + {3\over 2})+ m^2 c^4
-3 mc^2 \hbar w  \eqno(20) $$

\noindent In three dimensions (3+1), the Klein-Gordon oscillator
in a noncommutative space is given by

$$ c^2  [(\vec p +imw\vec r)\cdot (\vec p - i m w \vec r)] \star \psi =(E^2 -
m^2 c^4) \psi \eqno(21)$$

\noindent which is equivalent to the following equation in a
commutative space

$$ c^2 [\ \vec p + imw(\vec r + {{\vec \theta \times \vec p}\over {2 \hbar}})] \cdot
[ \ \vec p - i m w (\vec r + {{\vec \theta \times \vec p}\over {2
\hbar}})]\  \psi =(E^2 - m^2 c^4) \psi \eqno(22)$$

\noindent After straightforward calculation, we arrive at the
following equation ($\vec \theta = \theta \hat k $)

$$ c^2 [ (1+ {{m^2 w^2 \theta^2 }\over 4 \hbar^2})( p_x^2 + p_y^2 )  + \ m^2 w^2 (x^2 + y^2)-
 {{m^2 w^2 \theta}\over \hbar} \ L_z  ] \psi   $$

$$ =[(E^2 - m^2 c^4)-c^2(p_z^2 + m^2 w^2 z^2 - 3 m \hbar w  ) ]\psi \eqno(23)$$

\noindent  The above equation can be solved exactly and has a
similar behaviour to the dynamics of a scalar particle in a
constant magnetic field which is in the z direction and has a
coupling with an oscillator in the z direction. It should be
noted that a noncommutative space is not isotropic and the energy
eigenvalues are given by

$$ E_{n_xn_yn_zm_\ell}^2= 2mc^2\hbar w_1 (n_x+n_y+1)+2mc^2\hbar w (n_z+{1\over 2}) -
({m^2 w^2 \theta \over \hbar}) m_\ell \hbar +  m^2 c^4 - 3 m c^2
\hbar w   \eqno(24)$$

\noindent where

$$ w_1 = w {\sqrt{1+ {{m^2 w^2 \theta^2} \over {4 \hbar^2}}}}  \eqno(25) $$

\noindent It should be noted that the following map between the
old and new parameters in Eqs. (12) and (18) does not exist.

$$ c^2 (1+ {{m^2 w^2 \theta^2 }\over 4 \hbar^2}) \longrightarrow
c^2_{n} \eqno(26)$$

$$ c^2 m^2 w^2 \longrightarrow {{e^2_{n} B^2_{n}}\over 4  } \eqno(27)$$

$$ {{c^2 m^2 w^2 \theta}\over \hbar} \longrightarrow {e_{n} c_{n} B_{n} }  \eqno(28)$$

\vskip 0.2in \centerline{III. \bf \ The Dirac oscillator in a
noncommutative space } \vskip 0.1in

 The Dirac oscillator in a commutative space is defined
by the following substitution suggested by Ito et al. [15] see
also [16,17,18,19],

$$ {p_i}\rightarrow {p_i}-i\beta mw{x_i} \eqno(29) $$

$$ [ c \vec \alpha \cdot (\vec p -i\beta mw \vec r) + \beta mc^2] \psi(\vec r)= E \psi(\vec r)
\eqno(30) $$

 \noindent where

$$\psi(\vec r)=\pmatrix{\psi_a(\vec r)\cr \psi_b(\vec r) \cr}
  \, \ ,\  \vec \alpha=\pmatrix{0 & \vec \sigma \cr
 \vec \sigma & 0\cr} \ , \  \beta=\pmatrix{I & 0 \cr
 0& -I\cr}  \eqno(31) $$

\noindent A straightforward calculation leads to the following two
simultaneous equations

$$c \vec \sigma \cdot (\vec p + imw \vec r ) \psi_b(\vec
r)+mc^2  \psi_a(\vec r)= E \psi_a(\vec r)  \eqno(32)$$

$$ c \vec \sigma \cdot (\vec p - imw \vec r ) \psi_a(\vec
r)- mc^2 \psi_b(\vec r)= E \psi_b(\vec r)   \eqno(33)$$

\noindent In Eq. (9), $\psi_b(\vec r)$ is the small component of
the wave function, which tends to zero in the non-relativistic
limit. After some simple rearrangement and use of familiar
properties of the spin matrices we find

$$  c^2[\  p_x^2 + p_y^2   + m^2 w^2 (x^2 +
y^2)]\psi_a =[(E^2 -m^2 c^4 )- c^2(\  p_z^2 + m^2 w^2 z^2 - { 4 m
w \over \hbar}\vec S \cdot \vec L -3 m \hbar w )]\psi_a \eqno(34)
$$

\noindent and one may calculate the energy eigenvalues

$$ E^2=m^2 c^4 + (2n +l -2j +1)\hbar w mc^2  \ \ \ \ \ \ \ j=l + {1\over
2} \eqno(35) $$
$$ E^2=m^2 c^4 + (2n +l +2j + 3)\hbar w mc^2  \ \ \ \ \ \ \ j=l - {1\over
2} \eqno(36) $$

 \noindent The Dirac oscillator in a noncommutative space is given
 by the following equation,

$$ [ c \vec \alpha \cdot (\vec p -i\beta mw \vec r) + \beta
mc^2]\star \psi(\vec r)= E \psi(\vec r) \eqno(37)$$

\noindent  Using the new coordinates (5) in a commutative space,
we can map the Dirac oscillator in a noncommutative space to a
commutative one,

$$ [ c \vec \alpha \cdot (\vec p -i\beta mw (\vec r + {{\vec \theta \times \vec p}\over {2 \hbar}}))
+ \beta mc^2] \psi(\vec r)= E \psi(\vec r) \eqno(38)$$

 \noindent The final result for the $ \psi_a $
component is given by

$$  c^2[ (1+{m^2 w^2 \theta^2 \over 4 \hbar^2})(p_x^2 + p_y^2)  + m^2 w^2 (x^2 +
y^2) -  {m^2 w^2 \theta \over \hbar }(L_z + 2 S_z )]\psi_a = $$

$$[(E^2 -m^2 c^4 )- c^2( p_z^2 + m^2 w^2 z^2 - { 4  m w \over \hbar}\vec S \cdot
\vec L -3 m \hbar w ) - {2m w c^2 \over \hbar^2 } (\vec S \times
\vec p)\cdot (\vec \theta\times \vec p) ]\psi_a \eqno(39) $$

\noindent where $\theta_{ij}= \epsilon_{ijk}\theta_k $ and $
\theta_k = (0,0,\theta)$.

\noindent  The result shows that the Dirac oscillator in a
noncommutative space has a similar behaviour to the Dirac
equation in a commutative space describing  motion of a fermion
in a magnetic field along the z-axis, although an exact map
between the parameters does not exist.  The other terms are the
same as commutative space, i.e. a constant term, an oscillator in
the z direction and a spin-orbit coupling term; however, a new
interaction term appears which depends on the noncommutativity
parameter $\theta $, spin and also on the momentum operator.  The
last term has also an interpretation. For a charged particle in a
noncommutative space an electric dipole moment appears [22] which
is proportional to ${\vec \mu}_e \propto \vec \theta \times \vec
p $. The last term is similar to the spin orbit coupling in a
Hydrogen atom Hamiltonian and can be interpreted as a self
interaction. A moving particle with an electric dipole causes a
magnetic field which self-interacts with magnetic moment $\vec
\mu_m = ({e\over mc})\vec S $ of the particle

$$(\vec \theta\times \vec p)\cdot (\vec S \times
\vec p)   \propto \vec \mu_e \cdot (\vec S \times \vec p)$$

$$ \ \ \ \ \ \ \ \ \ \ \ \ \ \  \ \ \ \ \ \ \ \ \ \  \propto
{ \vec \mu}_m \cdot (\vec \mu_e \times \vec p)  \eqno(40)$$

\noindent It is interesting that without any field theory
calculations, the Dirac oscillator in a noncommutative space
implies an electric dipole moment for a charged particle. The
energy eigenvalues can be calculated exactly if we do not
consider the last term. The eigenvalues are given by

$$ E_{n_x n_y n_z m_\ell m_s }^2= 2mc^2 \hbar w_1 (n_x+n_y+1)+ 2mc^2\hbar w (n_z+ {1\over
2}) - {m^2 w^2 c^2 \theta }( m_{\ell} + 2 m_s)$$

$$\ \ \ \ \ \ \ \ \ \ \ \ \ \
- {2 m c^2 \hbar w }[j(j+1)-\ell (\ell +1)-s(s+1)] +
 m^2 c^4 - 3mc^2 \hbar w \eqno(41)$$

 \noindent The energy eigenvalues indicate a similarity to the {\it anomalous Zeeman effect}.

 \noindent In two dimensions $(2+1)$, this problem is exactly solvable.
 The same procedure can be applied for the Dirac equation in 2+1
dimensions. In this case the Dirac oscillator can be written as

$$ [ c \vec \alpha \cdot (\vec p -imw\beta \vec r ) + \beta m c^2
]\psi=E \psi \eqno(42)$$

\noindent where $ \vec \alpha $ and $ \beta $ are the Pauli
matrices and the wave function  is a two component spinor
$(\psi_A, \psi_B) $ .

$$  \alpha_x = \pmatrix{0 & 1 \cr
1  & 0\cr} \  \  \  \ \ \  \ \alpha_y = \pmatrix{0 & - i \cr i  &
0\cr} \ \ \ \ \ \ \ \beta =\pmatrix{1 & 0  \cr 0 & -1 \cr}
\eqno(43)$$

\noindent The two dimensional Dirac equation can be separated to
two equations

$$ c^2[(p_x^2 + p_y^2) +  m^2 w^2 (x^2 + y^2)  - {4 m w\over \hbar}\ ({\hbar \over 2})
 L_z - 2 m \hbar w ]\psi_A = (E^2
- m^2 c^4) \psi_A \eqno(44) $$

$$ c^2[(p_x^2 + p_y^2) +  m^2 w^2 (x^2 + y^2) + {4 m w\over \hbar}\ ({-\hbar \over 2})
L_z+ 2 m \hbar w ]\psi_B = (E^2 - m^2 c^4) \psi_B \eqno(45) $$

\noindent These equations are similar to the three dimensional
case and is equivalent to a two dimensional relativistic
oscillator with additional spin-orbit terms and a constant of
energy which has a different sign for particles and
antiparticles. It should be noted that in three dimensions the
spin-orbit term has a different sign for particles and
antiparticles, but in the case of two dimensions, as the sign of
spin also appears in the equation, it has the same sign in Eqs.
(44) and (45). In a noncommutative space, the two dimensional
Dirac oscillator can be written as

$$ c[(p_x -i p_y )+ i m w [(x- {\theta p_y \over 2 \hbar})-i(y+{\theta p_x\over 2\hbar}) ]]
\psi_B= (E-mc^2) \psi_A \eqno(46)$$

$$ c[(p_x + i p_y ) - i m w [(x- {\theta p_y \over 2 \hbar})+ i(y+{\theta p_x\over 2\hbar}) ]]
\psi_A= (E+mc^2) \psi_B \eqno(47)$$

\noindent After straightforward calculation we arrive at the
following equations for a particle with spin $\hbar \over 2 $

$$ c^2[(1+ {m^2 w^2 \theta^2\over 4 \hbar^2 })(p_x^2+p_y^2)+ m^2 w^2 (x^2+y^2)
-{m^2 w^2\theta \over \hbar}(L_z + 2({\hbar \over 2}))]\psi_A= $$

$$ [(E^2-m^2 c^4) + c^2(2m \hbar w + {4mw\over \hbar} ({\hbar \over 2})L_z )
- {2 m w c^2 \over \hbar^2 }\  \theta ({\hbar \over 2})
(p_x^2+p_y^2) ]\psi_A \eqno(48)$$

\noindent and an antiparticle with spin $-\hbar \over 2 $

$$ c^2[(1+ {m^2 w^2 \theta^2\over 4 \hbar^2 })(p_x^2+p_y^2)+ m^2 w^2 (x^2+y^2)
-{m^2 w^2\theta \over \hbar}(L_z + 2({-\hbar \over 2}))]\psi_B= $$

$$ [(E^2-m^2 c^4) - c^2(2m\hbar w + {4mw\over \hbar} ({-\hbar \over 2})L_z
) + {2 m w c^2 \over \hbar^2 }\ \theta ({-\hbar \over 2})
(p_x^2+p_y^2) ]\psi_B \eqno(49)$$

\noindent The above equations are similar to the equation of
motion for a fermion on a plane and in a constant magnetic field;
however, the last term is an additional interaction which appears
in this case and is the same as the exotic term in a three
dimensional space. In two dimensions, the problem is exactly
solvable and the energy eigenvalues for Eq.(48) are given by

$$ E_{n_x n_y m_\ell}^2 = 2mc^2\hbar w_1(n_x+n_y+1)-m^2 w^2 \theta (m_\ell
\pm 1)- 2m c^2  \hbar w \ m_\ell + m^2 c^4 \mp 2 m c^2 \hbar w
\eqno(50)$$

\noindent where
$$w_1 = 1 \pm {m w \theta \over \hbar} \eqno(51) $$

\noindent  The critical values of $\theta = \mp {\hbar \over m w}$
can be interpreted as a point which a resonance occurs. Such
resonance points are usually appear in dynamics of a particle with
spin and in a magnetic field.

\vskip 0.2in \centerline{IV. \bf \ Conclusion } \vskip 0.1in

We conclude that the known similarity between an oscillator in a
noncommutative space and a particle in a constant magnetic field
[10,11,13,14,20,21] can be extended to a relativistic motion. The
problem is exactly solvable in the spinless cases. However, for
the particles with spin or for the Dirac Oscillator in a
noncommutative space a new term in the Hamiltonian will appear
which can be interpreted as a self-interaction for a charged
particle with a dipole electric  and magnetic moments.

\vskip 0.2in \centerline{\bf \ \ Acknwoledgements} \vskip 0.2in
Our thanks go to the Isfahan University of Technology and
Institute for Studies in Theoretical Physics and Mathematics for
their financial support.

\vskip 0.2in
\centerline{\bf \ \  References} \vskip 0.1in

\noindent [1] \ M. R. Douglas and N. A. Nekrasov, Rev. Mod. Phys.
{\bf 73}(2001)977-1029.

\noindent \ \ \ \ \  hep-th/0106048.

\noindent [2] \ A. Connes, M. R. Douglas and A.Schwarz, JHEP {\bf
9802} (1998) 003, hep-th/9808042.

\noindent [3] \ N. Seiberg and E. Witten, JHEP {\bf 9909} (1999)
032. hep-th/9908142.

\noindent [4] \ L. Susskind, hep-th/0101029.

\noindent [5] \ M. Chaichian, A. Demichev, P. Presnajder, Nucl.
Phys. {\bf B567} 360 (2000),

\noindent \ \ \ \ \  hep-th/9812180.

\noindent [6] \ L. Alvarez-Gaume, S.R. Wadia, Phys. Lett. {\bf
B501} 319 (2001), hep-th/0006219.

\noindent [7] \ L. Mezincescu, hep-th/0007046.

\noindent [8] \ V. P.Nair, Phys. Lett. {\bf B505 } (2001) 249,
hep-th/0008027.

\noindent [9] \ M. Chaichian et al, Phys. Rev. Lett. {\bf
86}(2001) 2716, hep-th/0010175.

\noindent [10] \ J. Gamboa, M. Loewe and J. C. Rojas, Phys. Rev.
{\bf D64} (2001) 067901,

\noindent \ \ \ \ \ \  hep-th/0010220.

\noindent [11] \ S. Belluchi et al, Phys. Lett. {\bf B522} (2001)
345, hep-th/0106138.

\noindent [12] \ C. Acatrinei, JHEP {\bf 0109} (2001) 007,
hep-th/0107078.

\noindent [13] \ J. Gamboa et al, Int. J. Mod. Phys. {\bf A17}
(2002) 2555-2566, hep-th/0106125.

\noindent [14] \ J. Gamboa et al, Mod. Phys. Lett. {\bf A16}
(2001) 2075-2078, hep-th/0104224.

\noindent [15] \ D. Ito, K. Mori and E. Carrieri. (1967). Nuovo
Cimento {\bf 51A}, 1119.

\noindent [16] \ M. Moshinsky and A. Szczepaniak, J. Phys. A.
Math. Gen. {\bf 22} (1989) L817-L819.

\noindent [17] \ M. Moreno and A. Zentella, J. Phys. A: Math.
Gen. {\bf 22} (1989) L821-L825.

\noindent [18] \ J. Benitez et al, Phys. Rev. Lett. {\bf 64}
(1990) 14-1643.

\noindent [19] \ P. Strange, Relativistic Quantum Mechanics,
(1998) Camb. Univ. Press.

\noindent [20] \ B. Muthukumar and P. Mitra,  Phys.Rev. {\bf D66}
(2002) 027701, hep-th/0204149.

\noindent [21] \  V. M. Villalba, A. Rincon Maggiolo, Eur.Phys.J.
{\bf B22  }  (2001) 31, cond-mat/0107529.

\noindent [22] \ I. F. Riad and M.M. Sheikh-Jabbari,  JHEP 0008
(2000) 045, hep-th/0008182.

 \vfill\eject

\bye